# Low-Cost Sensor Fusion Framework for Organic Substance Classification and Quality Control Using Classification Methods

Borhan Uddin Chowdhury
*Ingram School of Engineering*
*Texas State University*
San Marcos, USA
borhan.me@txstate.edu

Damian Valles
*Ingram School of Engineering*
*Texas State University*
San Marcos, USA
dvalles@txstate.edu

Md Raf E Ul Shougat
*Ingram School of Engineering*
*Texas State University*
San Marcos, USA
shougat@txstate.edu

*Abstract*—We present a sensor-fusion framework for rapid, non-destructive classification and quality control of organic substances, built on a standard Arduino Mega 2560 microcontroller platform equipped with three commercial environmental and gas sensors. All data used in this study were generated in-house: sensor outputs for ten distinct classes—including fresh and expired samples of apple juice, onion, garlic, and ginger, as well as cinnamon and cardamom—were systematically collected and labeled using this hardware setup, resulting in a unique, application-specific dataset. Correlation analysis was employed as part of the preprocessing pipeline for feature selection. After preprocessing and dimensionality reduction (PCA/LDA), multiple supervised learning models—including Support Vector Machine (SVM), Decision Tree (DT), and Random Forest (RF), each with hyperparameter tuning, as well as an Artificial Neural Network (ANN) and an ensemble voting classifier—were trained and cross-validated on the collected dataset. The best-performing models, including tuned Random Forest, ensemble, and ANN, achieved test accuracies in the 93–94% range. These results demonstrate that low-cost, multisensory platforms based on the Arduino Mega 2560, combined with advanced machine learning and correlation-driven feature engineering, enable reliable identification and quality control of organic compounds.

## I. Introduction

Accurate and non-destructive identification of organic materials is crucial for quality control, food safety, and environmental monitoring across various industries. Traditional methods—such as laboratory chemical analysis or subjective visual inspection—are often costly, time-consuming, or prone to error, especially when distinguishing between fresh and expired products or similar organic substances. Recent advances in sensor technology and machine learning have enabled new approaches that combine multiple sensor modalities to improve reliability and automation in substance classification.

Sensor fusion, which integrates data from diverse sources such as environmental and gas sensors, has emerged as a powerful strategy to capture the complex signatures of organic compounds. Prior research has demonstrated that combining signals from temperature, humidity, pressure, and gas concentrations can enhance detection accuracy in applications ranging from food spoilage detection to industrial safety monitoring. However, many existing studies rely on expensive hardware, proprietary platforms, or limited datasets, making them less accessible for widespread adoption.

This study addresses these challenges by developing a low-cost, open-source sensor fusion platform based on the Arduino Mega 2560 and three commercial sensor modules. The system collects synchronized measurements of temperature, humidity, pressure, and multiple gas concentrations ($CO_2$, $NO_2$, VOCs, ethanol, CO), providing a rich feature set for classification. Using this setup, we generated a unique dataset that encompasses ten classes of organic substances: fresh and expired samples of apple juice, onion, garlic, and ginger, as well as cinnamon and cardamom, all collected under real-world conditions.

The collected data were systematically processed through scaling and correlation-driven feature selection to ensure model robustness. Dimensionality reduction techniques—including principal component analysis (PCA), linear discriminant analysis (LDA), t-distributed stochastic neighbor embedding (t-SNE), and uniform manifold approximation and projection (UMAP)—were employed for exploratory visualization and to assess class separability. Multiple supervised learning models, including tuned Decision Tree, Random Forest, Support Vector Machine, Artificial Neural Network (ANN), and an ensemble voting classifier, were trained and evaluated using stratified cross-validation and comprehensive performance metrics.

Our results show that the best-performing models achieve test accuracies in the 93–94% range, demonstrating that affordable sensor fusion, combined with advanced machine learning, enables reliable and scalable classification of organic substances. This approach offers a practical solution for rapid, non-destructive quality control and safety monitoring, with potential applications in resource-limited settings.

## II. Background

The detection and classification of volatile organic compounds (VOCs) have become increasingly critical in environmental monitoring, food safety, and industrial applications. Among various approaches, electronic nose (e-nose) systems have emerged as a low-cost, portable, and intelligent alternative to conventional analytical instruments. These systems mimic the human olfactory system by employing gas sensor arrays and pattern recognition models to identify VOCs [1][2]. Tin oxide ($SnO_2$)-based metal oxide semiconductor (MOS) sensors are particularly popular due to their low cost, ease of integration, and reasonable selectivity toward gases such as alcohols, ketones, and hydrocarbons [3][4]. With the advancement of miniaturized computing platforms such as Arduino and Raspberry Pi, many studies have successfully implemented e-nose systems for food quality detection and industrial gas monitoring using compact sensor arrays [5][6].

Integrating machine learning (ML) and deep learning (DL) techniques with gas sensor systems has significantly improved VOC classification accuracy, robustness to sensor drift, and generalization across environments. Algorithms such as support vector machines (SVM), artificial neural networks (ANN), convolutional neural networks (CNN), and long short-term memory (LSTM) networks have been successfully used to model time-series data from e-nose signals [7][8]. For example, hybrid CNN-LSTM architectures have outperformed standalone models in classifying complex odor profiles in spirit samples and industrial VOC mixtures [9], [10]. Studies have also explored early and late fusion models that combine heterogeneous sensor modalities or multiple time steps, thereby enhancing multi-class discrimination capability [11][12]. When deployed on embedded platforms, these techniques balance inference efficiency and classification performance, making them ideal for real-world, low-power applications like the present study.

Recent works have also emphasized multimodal sensor fusion, employing surface acoustic wave (SAW) sensors, electrochemical sensors, or e-tongues alongside MOS arrays to improve sensitivity and selectivity [13][14]. Others have incorporated transfer learning and adaptive calibration techniques to handle sensor drift over time [15][16]. While some studies focus on large-scale industrial datasets, others demonstrate effective classification of 5–12 VOC classes using fewer than 1,000 samples, highlighting the feasibility of using small-scale datasets for embedded AI [17][18]. The present work builds on this literature by developing an Arduino-compatible sensor fusion model capable of distinguishing ten VOC classes using multiple supervised learning models—including Support Vector Machine (SVM), Decision Tree (DT), Random Forest (RF), Artificial Neural Networks (ANN), and an ensemble voting classifier. While traditional ML models were optimized through hyperparameter tuning, the ANN models were explored across multiple structural configurations. This approach offers a compact, low-cost, and scalable solution for gas classification in resource-constrained environments.

### III. Data

#### A. Data Collection

Data was collected using an Arduino Mega 2560 equipped with three sensors: the Grove Multichannel Gas Sensor V2, Grove SGP30, and Grove BME680. These sensors measured various environmental and chemical parameters, including CO, $NO_2$, VOCs, ethanol, $CO_2$, TVOC, temperature, humidity, and pressure. The dataset comprised ten classes—fresh and expired variants of apple juice, onion, garlic, and ginger, along with cinnamon and cardamom. For each class, 10,000 samples were recorded. Sensor readings were collected over multiple runs and stored in individual CSV files, each containing all feature columns and a class label. These files were subsequently merged into a single master dataset, resulting in 100,000 samples. The dataset was balanced and underwent quality assurance through sensor calibration, stabilization before logging, and preprocessing techniques to address noise and potential anomalies. Fig. 1 shows the overall sensor integration and data acquisition workflow used to generate the dataset.

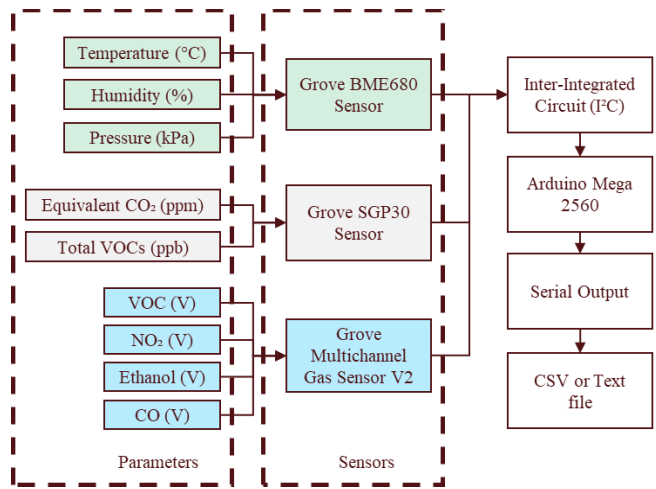

**Fig. 1.** Sensor fusion and data collection flow using Arduino Mega 2560.

#### B. Data Preprocessing

The raw sensor data for each organic substance was combined into a single dataset, with the target column labeled according to the substance name (e.g., "onion," "garlic," etc.). Only the target column was categorical; all other features were numeric, and there were no missing values, as the data was collected after allowing sensors to stabilize.

#### C. Feature Selection

Correlation analysis showed that the pressure feature had the highest positive correlation with the target variable, while temperature was strongly negatively correlated. Although this suggests that both features may be informative, they likely reflect uncontrolled environmental drift rather than meaningful chemical variation. Two dataset versions were compared to validate their impact: V1, including all features, and V2, excluding temperature and pressure. Model performance dropped slightly (~0.1 accuracy) after removing these features. To avoid overfitting to ambient conditions and ensure generalizable learning, V2 was used for all subsequent analyses.

#### D. Encoding and Scaling

The categorical target column was label-encoded to numeric values for machine learning compatibility. All feature columns were standardized using z-score normalization (StandardScaler) to ensure comparability across different measurement units.

#### E. Splitting

The dataset was first shuffled and split into training and test sets using stratified sampling to preserve class distribution. Subsequently, stratified k-fold cross-validation (with $k = 5$) was performed on the training set for model selection and hyperparameter tuning. This two-stage approach ensured a balanced representation of all classes during validation and final evaluation.

#### F. Dimensionality Reduction

Techniques like PCA, LDA, t-SNE, and UMAP were applied for exploratory visualization and to assess class separability. These steps ensured the dataset was clean, consistent, and focused on relevant features for organic substance classification.

## IV. Model Development

### A. Model Selection and Versions

Initially, three machine learning models—Support Vector Machine (SVM), Decision Tree (DT), and Random Forest (RF)—were selected for evaluation using four distinct feature set versions:

- **V1:** All available features included.
- **V2:** Excluded temperature and pressure features based on their minimal impact on model performance.
- **V3:** PCA dimensionality reduction applied to the V2 feature set.
- **V4:** LDA dimensionality reduction applied to the V2 feature set.

Fig. 2 illustrates the complete model development pipeline, including traditional ML workflows, feature set variations, hyperparameter tuning, ensemble construction, and ANN architecture exploration.

### B. Model Evaluation and Feature Selection

Performance analysis was conducted through stratified 5-fold cross-validation across all four dataset versions. While V1 (all features) yielded slightly higher accuracy, approximately 0.1% more than V2, the improvement was deemed negligible and potentially inflated due to ambient drift captured by temperature and pressure. Dimensionality reduction techniques (V3 and V4) also did not significantly enhance classification performance compared to the simplified V2 set. Therefore, V2 was selected for further model development and hyperparameter tuning, offering a more robust and generalizable feature set.

### C. Hyperparameter Tuning

GridSearchCV was utilized for comprehensive hyperparameter optimization exclusively on the V2 feature set, significantly improving accuracy compared to baseline (untuned) models. Optimal hyperparameters were identified through extensive grid searches.

### D. Ensemble Classifier Development

A majority-voting ensemble classifier was constructed using the optimized SVM, DT, and RF models from the V2 feature set. This ensemble aimed to enhance classification robustness by aggregating predictions from diverse algorithms.

### E. Artificial Neural Network (ANN)

In parallel, several ANN architectures were explored using sequential models:

- Baseline (initial sequential model)
- Deeper (increased number of hidden layers)
- Wider (increased number of neurons per hidden layer)
- L2 Regularization (applied regularization to reduce overfitting)
- RMSprop optimizer (alternative optimizer to Adam)

Architecture comparison indicated minimal improvements from structural changes, with the "wider" model slightly outperforming others. Given the negligible differences, the baseline architecture was considered near optimal.

The combined results from the tuned classical models, ensemble, and ANN underscored the robust performance achievable through sensor fusion and advanced machine-learning techniques.

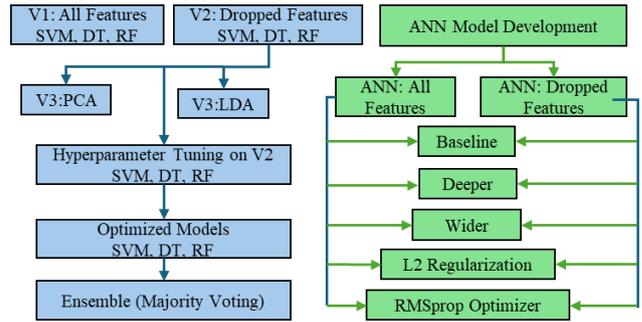

**Fig. 2.** Flowchart of Model Development and ANN Architecture Exploration.

In total, 26 distinct models were developed and evaluated, comprising twelve baseline traditional models across four dataset versions, three tuned classifiers on the selected V2 feature set, one ensemble model, and ten ANN variants (spanning five architectural configurations on both original and reduced feature sets).

## V. Results

### A. Correlation Analysis

Feature Correlation Analysis before applying dimensionality reduction and modeling, correlation analysis was performed to evaluate the relationship between each feature and the target variable. As shown in Fig. 3, pressure exhibited the highest positive correlation with the target, followed by CO and VOC1. Conversely, temperature showed a strong negative correlation.

While this may suggest that pressure and temperature are essential predictors, their correlation is likely due to uncontrolled environmental conditions during data collection. Since pressure and temperature were not consistently maintained across samples, their inclusion could lead to an inflated model performance by capturing ambient drift rather than substance-specific characteristics.

These two features were excluded in version V2 and beyond to avoid overfitting and ensure that classification is based on meaningful chemical signals rather than external environmental factors.

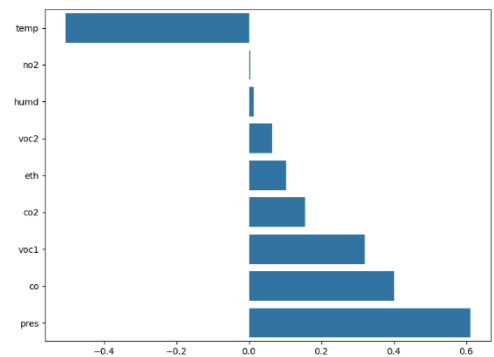

**Fig. 3.** Correlation of individual sensor features with the target class label.

### B. Visualizations

Multiple projection techniques were applied to the complete and reduced feature sets to assess class separability and the effects of dimensionality reduction. Fig. 4 shows the class-wise distributions using PCA and LDA, illustrating moderate separation between certain classes.

The t-SNE and UMAP plots in Fig. 5 further demonstrate that the remaining features are sufficient to form separable clusters even after dropping temperature and pressure. This supports the decision to exclude those environmentally sensitive variables, as their removal did not compromise the underlying structure of the data.

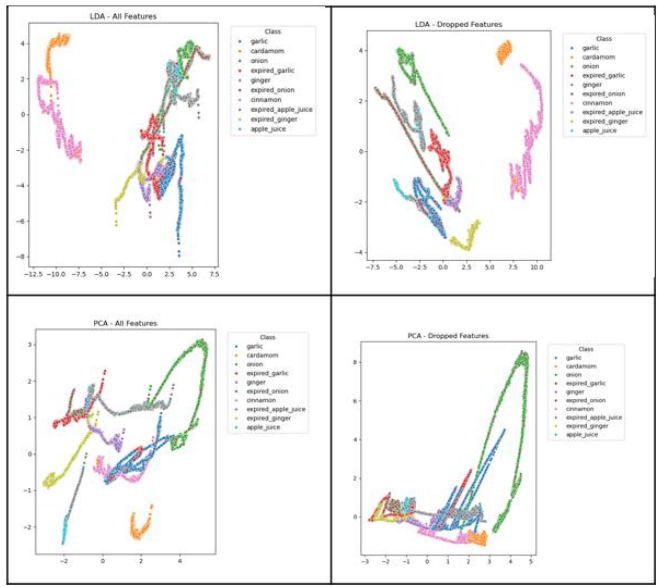

**Fig. 4.** LDA and PCA projection on the complete and reduced feature sets.

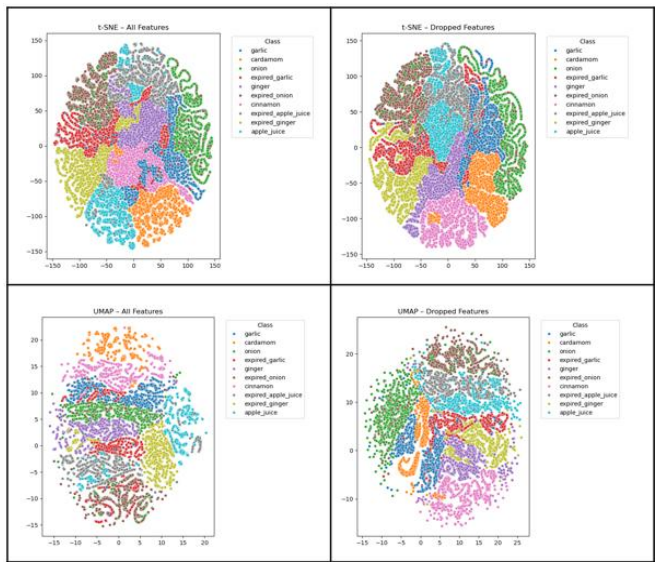

**Fig. 5.** t-SNE and UMAP projection on the complete and reduced feature sets.

### C. Cross-Validation Accuracy Comparison

Cross-validation accuracies were compared across four dataset versions (V1–V4) for each classifier (SVM, RF, DT) to evaluate the impact of feature engineering and dimensionality reduction. The results are presented in Fig. 6.

V1 and V2 outperformed the other versions, achieving the highest accuracy across all three models. While V1, which includes all features, consistently yielded slightly higher accuracy than V2, the difference was marginal, typically around 0.1%, and likely influenced by uncontrolled environmental variations. Since V2 excludes temperature and pressure, it avoids learning from potentially spurious patterns caused by ambient drift. Given the minimal performance trade-off, reduced complexity, and improved generalizability, V2 was selected as the final version for model tuning and further evaluation.

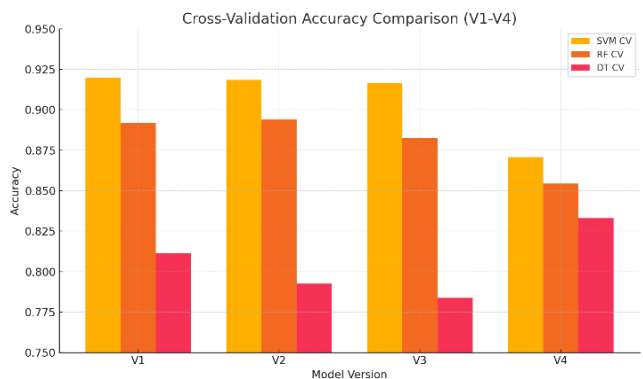

**Fig. 6.** Cross-validation accuracy comparison of SVM, Random Forest (RF), and Decision Tree (DT) across four dataset versions (V1–V4).

### D. Hyperparameter Tuning

To improve the baseline model's performance on the selected V2 feature set, hyperparameter tuning was performed using GridSearchCV with stratified 5-fold cross-validation. Parameter grids were defined separately for each classifier—SVM, Random Forest (RF), and Decision Tree (DT)—exploring different kernel types and regularization strengths for SVM, maximum depth and minimum leaf sizes for DT, and tree count and feature subset sizes for RF.

The best cross-validation and evaluation scores obtained after tuning were:

- SVM (CV): 0.9272, Train: 0.9304, Test: 0.9301
- RF (CV): 0.9425, Train: 0.9438, Test: 0.9427
- DT (CV): 0.9385, Train: 0.9397, Test: 0.9396

These improvements were particularly notable for RF and DT compared to their untuned counterparts. Fig. 7 compares cross-validation accuracies before and after tuning for all three classifiers to visualize the effectiveness of hyperparameter tuning. The improvement across models confirms the benefits of targeted tuning.

Additionally, learning curves for the tuned models were plotted to assess generalization performance. As shown in Fig. 8, training and validation curves converged steadily as the sample size increased, indicating low variance and minimal overfitting. All three models continued to improve with additional training data, suggesting capacity for further enhancement.

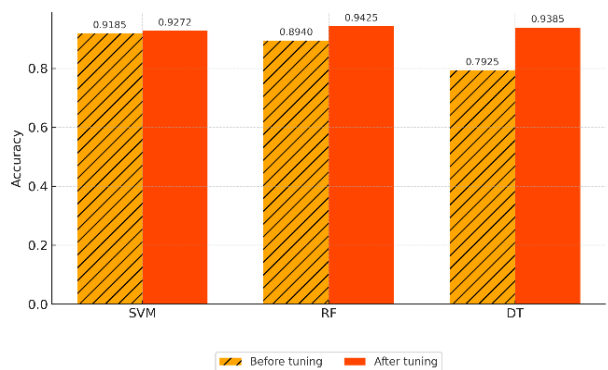

**Fig. 7.** CV accuracy before and after hyperparameter tuning for SVM, RF, and DT models.

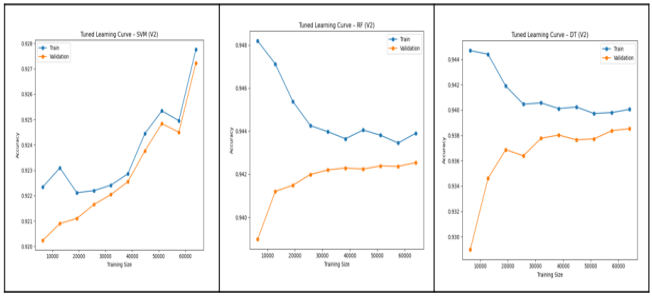

**Fig. 8.** Tuned learning curves for SVM, Random Forest (RF), and Decision Tree (DT) models on the V2 feature set.

These results validate both the effectiveness and stability of the optimized models.

*E. Ensemble Classifier Evaluation*

A soft-voting ensemble was constructed using the tuned SVM, RF, and DT models to improve classification robustness. Each model contributed probabilistic predictions, and the final class label was determined by averaging the predicted probabilities and selecting the class with the highest consensus. Fig. 9 shows the learning curve of the ensemble model. The ensemble demonstrates stable convergence, with validation accuracy closely tracking training performance as the dataset size increases.

The ensemble model achieved substantial and consistent results, with a training accuracy of 94.27%, a test accuracy of 94.15%, and a cross-validation accuracy of 94.15% ± 0.16%. These metrics indicate reliable performance across training, validation, and unseen test data. However, while strong and consistent, the ensemble did not surpass the tuned Random Forest, which had a slightly higher cross-validation accuracy of 94.25%. Fig. 10 shows the confusion matrix for test sets. These highlight the ensemble's effective handling of most classes with minimal confusion, particularly improving predictions for specific difficult-to-classify samples.

The classification report in Table I further supports this conclusion, with macro-averaged precision, recall, and F1 Scores all above 0.94, highlighting consistent and balanced class-wise performance. Finally, Fig. 11 displays the ensemble's ROC curve on the test set. It shows excellent discrimination, with a micro-average AUC of 1.00 and a macro-average AUC of 0.99, confirming the ensemble's capability across all classes.

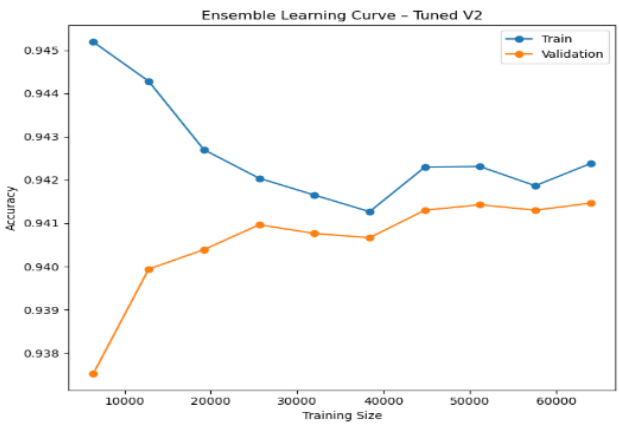

**Fig. 9.** Learning curve for the Ensemble model.

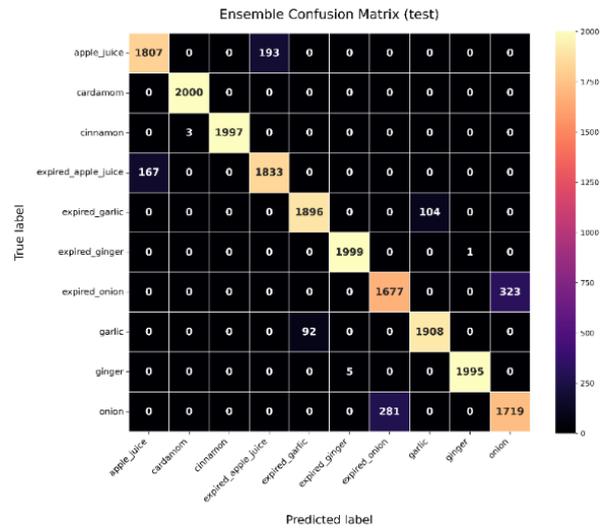

**Fig. 10.** Confusion matrix of the ensemble model on the test set.

TABLE I. CLASSIFICATION REPORT OF THE ENSEMBLE MODEL ON THE TEST SET

| class | precision | recall | f1-score | support |
|---|---|---|---|---|
| apple_juice | 0.9154 | 0.9035 | 0.9094 | 2000 |
| cardamom | 0.9985 | 1.0000 | 0.9993 | 2000 |
| cinnamon | 1.0000 | 0.9985 | 0.9992 | 2000 |
| expired_apple_juice | 0.9047 | 0.9165 | 0.9106 | 2000 |
| expired_garlic | 0.9537 | 0.9480 | 0.9509 | 2000 |
| expired_ginger | 0.9975 | 0.9995 | 0.9985 | 2000 |
| expired_onion | 0.8565 | 0.8385 | 0.8474 | 2000 |
| garlic | 0.9483 | 0.9540 | 0.9511 | 2000 |
| ginger | 0.9995 | 0.9975 | 0.9985 | 2000 |
| onion | 0.8418 | 0.8595 | 0.8506 | 2000 |
| accuracy | 0.9416 | 0.9416 | 0.9416 | 0.9416 |
| macro avg | 0.9416 | 0.9416 | 0.9415 | 20000 |
| weighted avg | 0.9416 | 0.9416 | 0.9415 | 20000 |

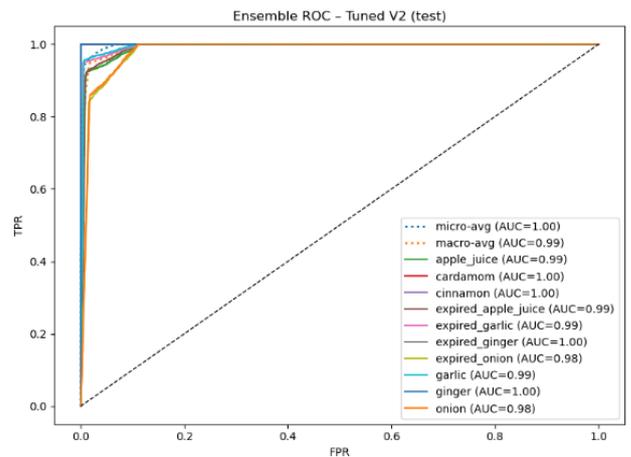

**Fig. 11.** ROC curve for the tuned ensemble model on the test set.

*F. Tuned Random Forest Evaluation*

After hyperparameter tuning, the Random Forest (RF) model emerged as the top-performing individual classifier. It consistently achieved the highest scores across all evaluation metrics:

- Cross-Validation Accuracy: 94.25%
- Train Accuracy: 94.38%
- Test Accuracy: 94.27%

These results underscore RF's ability to generalize well while maintaining a low variance between training and test performance. Fig. 12 presents the confusion matrix for the test set. The model demonstrates excellent discrimination, correctly classifying most instances with minimal class confusion.

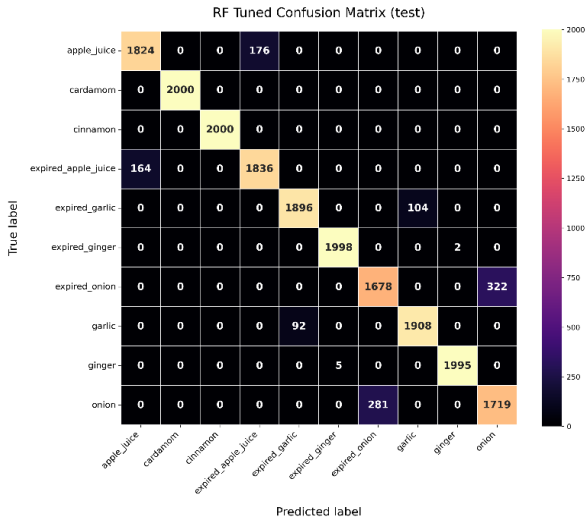

**Fig. 12.** Confusion matrix of the tuned Random Forest on the test set.

Table II displays the classification report test set, confirming high macro-averaged precision, recall, and F1-scores above 0.94.

TABLE II. CLASSIFICATION REPORT OF THE TUNED RANDOM FOREST MODEL ON THE TEST SET

| class | precision | recall | f1-score | support |
|---|---|---|---|---|
| apple_juice | 0.9175 | 0.9120 | 0.9147 | 2000 |
| cardamom | 1.0000 | 1.0000 | 1.0000 | 2000 |
| cinnamon | 1.0000 | 1.0000 | 1.0000 | 2000 |
| expired_apple_juice | 0.9125 | 0.9180 | 0.9153 | 2000 |
| expired_garlic | 0.9537 | 0.9480 | 0.9509 | 2000 |
| expired_ginger | 0.9975 | 0.9990 | 0.9983 | 2000 |
| expired_onion | 0.8566 | 0.8390 | 0.8477 | 2000 |
| garlic | 0.9483 | 0.9540 | 0.9511 | 2000 |
| ginger | 0.9990 | 0.9975 | 0.9982 | 2000 |
| onion | 0.8422 | 0.8595 | 0.8508 | 2000 |
| accuracy | 0.9427 | 0.9427 | 0.9427 | 0.9427 |
| macro avg | 0.9427 | 0.9427 | 0.9427 | 20000 |
| weighted avg | 0.9427 | 0.9427 | 0.9427 | 20000 |

Fig. 13 shows the ROC curve for the tuned RF model, which exhibits near-perfect separation with micro- and macro-average AUC values approaching 1.00. The tuned Random Forest is the strongest candidate based on accuracy, generalization, and class-wise performance. While it outperformed the ensemble model across key metrics, further evaluation is required before final model selection. The following section will explore and compare the performance of various ANN architectures to tuned Random Forest and ensemble benchmarks.

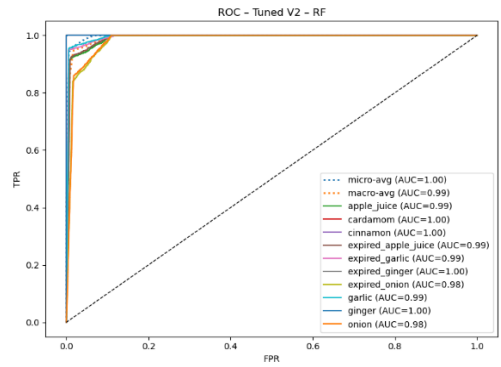

**Fig. 13.** ROC curve for the Random Forest model on the test set.

### G. Artificial Neural Network (ANN) Evaluation

To identify the most effective ANN design, five variants were evaluated on all features, including the dropped feature set (excluding temperature and pressure)

- Baseline: Standard sequential model with three dense layers
- Deeper: Increased number of hidden layers
- Wider: Increased neurons per layer
- L2 Regularization: Added weight penalties to reduce overfitting
- RMSprop Optimizer: Switched from Adam to RMSprop

Fig. 14 and Fig. 15 present train, validation, and test accuracies for all five ANN architectures on the complete and reduced feature sets, respectively.

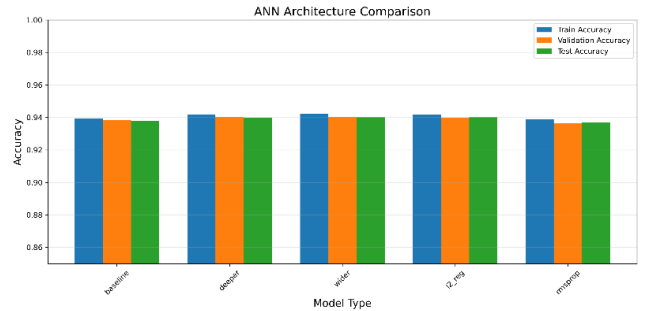

**Fig. 14.** Train, validation, and test accuracies of five ANN variants using all features.

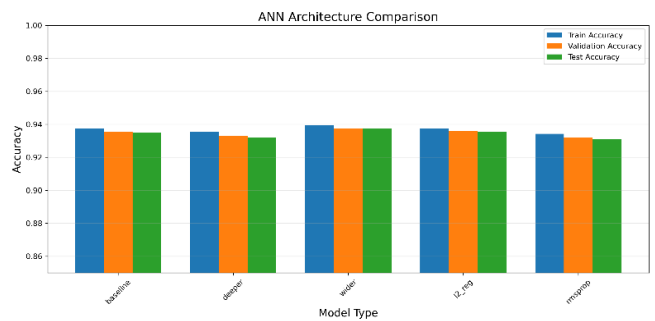

**Fig. 15.** Train, validation, and test accuracy of five ANN variants with temperature and pressure dropped.

When using all available features, the accuracy across all ANN architectures was tightly clustered around 94.00%, reflecting consistent and strong performance. Among the tested configurations, the wider architecture achieved the highest test accuracy of 94.01%, marginally outperforming the other variants. Furthermore, all models demonstrated a good balance between training, validation, and test accuracies, indicating stable generalization without overfitting.

The same architectural variants were re-evaluated after removing temperature and pressure from the input set. The trends remained largely consistent—the wider model again showed the best performance. However, the highest test accuracy dropped slightly to 93.73%, suggesting a minor decrease in model performance without the environmental features.

Overall, dropping temperature and pressure had a minimal impact on ANN accuracy. This outcome reinforces the observation in classical models: the ANN's predictive power is primarily driven by the core gas sensing features. Thus, ANN performance remains robust and generalizable even without these environmental variables.

The best-performing ANN configuration used the reduced feature set (dropping temperature and pressure) and a wider architecture with three hidden layers of sizes 512, 256, and 128, respectively, followed by batch normalization and dropout. The model included a GaussianNoise layer for input regularization and a final SoftMax output layer for 10-class classification, totaling approximately 173K parameters.

This configuration achieved a training accuracy of 93.94%, a validation accuracy of 93.74%, and a test accuracy of 93.73%. Fig. 16 shows the training loss and accuracy curves, confirming smooth convergence without overfitting. The Confusion matrix in Fig. 17 showed strong and consistent class-wise accuracy for the final ANN model. Table III provides the full classification report.

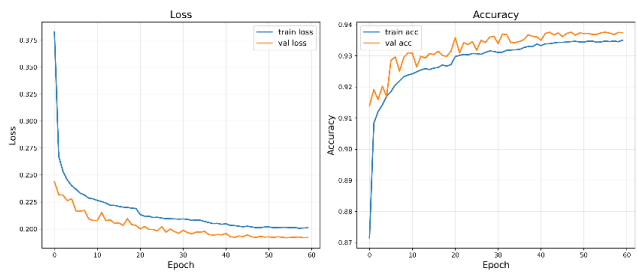

**Fig. 16.** Loss and accuracy curves of the wider ANN model with dropped features.

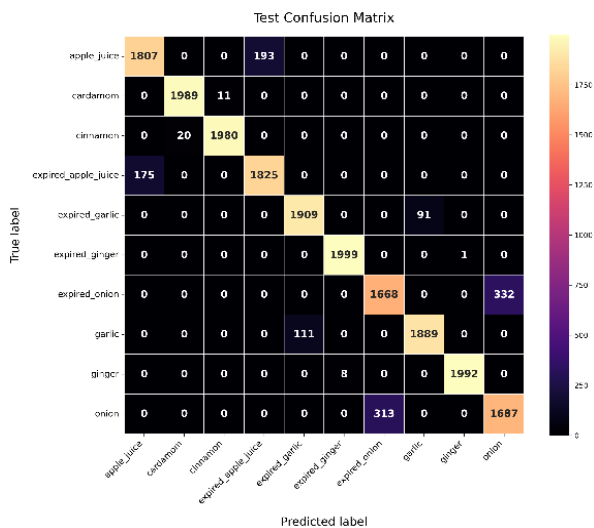

**Fig. 17.** Confusion matrix of the wider ANN model with dropped features on the test set.

Table III demonstrates that the macro-averaged precision, recall, and F1-score were all around 0.94.

TABLE III. CLASSIFICATION REPORT OF THE WIDER ANN MODEL WITH DROPPED FEATURES ON TEST SET

| class | precision | recall | f1-score | support |
|---|---|---|---|---|
| apple_juice | 0.9117 | 0.9035 | 0.9076 | 2000 |
| cardamom | 0.9900 | 0.9945 | 0.9923 | 2000 |
| cinnamon | 0.9945 | 0.9900 | 0.9922 | 2000 |
| expired_apple_juice | 0.9044 | 0.9125 | 0.9084 | 2000 |
| expired_garlic | 0.9450 | 0.9545 | 0.9498 | 2000 |
| expired_ginger | 0.9960 | 0.9995 | 0.9978 | 2000 |
| expired_onion | 0.8420 | 0.8340 | 0.8380 | 2000 |
| garlic | 0.9540 | 0.9445 | 0.9492 | 2000 |
| ginger | 0.9995 | 0.9960 | 0.9977 | 2000 |
| onion | 0.8356 | 0.8435 | 0.8395 | 2000 |
| accuracy | | | 0.9373 | 20000 |
| macro avg | 0.9373 | 0.9373 | 0.9372 | 20000 |
| weighted avg | 0.9373 | 0.9373 | 0.9372 | 20000 |

Despite this strong performance, the ANN model slightly underperformed the tuned Random Forest, which achieved a test accuracy of 94.27%. However, the ANN's performance was comparable, and generalization was robust, reinforcing that the ANN is a viable alternative classification method in sensor fusion.

## VI. CONCLUSION

This project demonstrates that a low-cost sensor fusion system, built around Arduino Mega 2560 and commercial gas sensors, provides a highly effective and affordable solution for rapid, non-destructive quality control of organic substances. The tuned Random Forest (RF) classifier was selected as the best-performing approach through rigorous model development and evaluation. It achieved a test accuracy of 94.27%, outperforming other models such as SVM, Decision Tree, ANN, and even the ensemble method. The tuned RF model delivered the highest accuracy and showed excellent generalization and balanced class-wise performance, reliably distinguishing between fresh and expired substances across multiple organic classes.

This approach is very cheap compared to traditional laboratory methods, leveraging affordable hardware and open-source tools. The system enables reliable, on-site identification of organic substance quality, including spoilage detection, making advanced quality control accessible even for small producers or resource-limited settings.

## VII. FUTURE WORK

Collecting more data and performing deeper hyperparameter tuning for future work can further improve results. Additionally, maintaining consistent temperature and pressure conditions during data collection will allow these features to be reliably integrated into the model, further enhancing classification performance.

REFERENCES

[1] A. D. Wilson and M. Baietto, "Applications and advances in electronic-nose technologies," *Sensors*, vol. 9, no. 7, pp. 5099–5148, 2009.
[2] H. Yang *et al.*, "A comprehensive review of VOCs as a key indicator in food authentication," *eFood*, vol. 6, no. 3, p. e70057, 2025.

[3] A. K. Srivastava, "Detection of volatile organic compounds (VOCs) using $SnO_2$ gas-sensor array and artificial neural network," *Sens. Actuators B Chem.*, vol. 96, no. 1–2, pp. 24–37, 2003.

[4] R. Dutta, E. L. Hines, J. W. Gardner, K. R. Kashwan, and M. Bhuyan, "Tea quality prediction using a tin oxide-based electronic nose: an artificial intelligence approach," *Sens. Actuators B Chem.*, vol. 94, no. 2, pp. 228–237, 2003.

[5] M. Wang and Y. Chen, "Electronic nose and its application in the food industry: a review," *Eur. Food Res. Technol.*, vol. 250, no. 1, pp. 21–67, 2024.

[6] L. Cheng, Q.-H. Meng, A. J. Lilienthal, and P.-F. Qi, "Development of compact electronic noses: A review," *Meas. Sci. Technol.*, vol. 32, no. 6, p. 062002, 2021.

[7] C. Bilgera, A. Yamamoto, M. Sawano, H. Matsukura, and H. Ishida, "Application of convolutional long short-term memory neural networks to signals collected from a sensor network for autonomous gas source localization in outdoor environments," *Sensors*, vol. 18, no. 12, p. 4484, 2018.

[8] J.-T. Sun and C.-H. Lee, "AI-driven sensor array electronic nose system for authenticating and recognizing aromas in spirit samples," *Sensors and Mater.*, vol. 37, no. 1, pp. 23–40, 2025.

[9] X. Pan, H. Zhang, W. Ye, A. Bermak, and X. Zhao, "A fast and robust gas recognition algorithm based on hybrid convolutional and recurrent neural network," *IEEE Access*, vol. 7, pp. 100954–100963, 2019.

[10] Q. Liu, X. Hu, M. Ye, X. Cheng, and F. Li, "Gas recognition under sensor drift by using deep learning," *Int. J. Intell. Syst.*, vol. 30, no. 8, pp. 907–922, 2015.

[11] P. Narkhede, R. Walambe, S. Mandaokar, P. Chandel, K. Kotecha, and G. Ghinea, "Gas detection and identification using multimodal artificial intelligence-based sensor fusion," *Appl. Syst. Innov.*, vol. 4, no. 1, p. 3, 2021.

[12] Y. Luo, W. Ye, X. Zhao, X. Pan, and Y. Cao, "Classification of data from electronic nose using gradient tree boosting algorithm," *Sensors*, vol. 17, no. 10, p. 2376, 2017.

[13] C. Li, "Sensor fusion models for integrating electronic nose and surface acoustic wave sensor for apple quality evaluation," unpublished, 2007.

[14] Z. Ye, Y. Liu, and Q. Li, "Recent progress in smart electronic nose technologies enabled with machine learning methods," *Sensors*, vol. 21, no. 22, p. 7620, 2021.

[15] R. Calvini and L. Pigani, "Toward the development of combined artificial sensing systems for food quality evaluation: A review on the application of data fusion of electronic noses, electronic tongues and electronic eyes," *Sensors*, vol. 22, no. 2, p. 577, 2022.

[16] H. W. Noh, Y. Jang, H. D. Park, D. Kim, J. H. Choi, and C.-G. Ahn, "A selective feature optimized multi-sensor based e-nose system detecting illegal drugs validated in diverse laboratory conditions," *Sens. Actuators B Chem.*, vol. 390, p. 133965, 2023.

[17] H. Anwar, T. Anwar, and S. Murtaza, "Review on food quality assessment using machine learning and electronic nose system," *Biosens. Bioelectron. X*, vol. 14, p. 100365, 2023.

[18] Y. Li, X. Huang, E. Witherspoon, Z. Wang, P. Dong, and Q. Li, "Intelligent electrochemical sensors for precise identification of volatile organic compounds enabled by neural network analysis," *IEEE Sens. J.*, 2024.